\documentclass[conference]{IEEEtran}
 \usepackage{fancyhdr}
\IEEEoverridecommandlockouts
\usepackage{cite}
\usepackage{amsmath,amssymb,amsfonts}
\usepackage{pifont}
\usepackage{algpseudocode}
\usepackage{graphicx}
\usepackage{textcomp}
\usepackage{kotex}
\usepackage{xcolor}
\usepackage{amsthm}
\usepackage{amsmath}
\usepackage{subfigure}
\usepackage[linesnumbered,ruled,vlined]{algorithm2e}
\usepackage{algorithmicx}
\usepackage{hyperref}
\usepackage{array}

\usepackage{booktabs}
\usepackage{tabularx}
\usepackage{authblk}
\usepackage{multicol}

\newcommand\cheolho[1]{{\color{black}#1}}

\def\BibTeX{{\rm B\kern-.05em{\sc i\kern-.025em b}\kern-.08em
    T\kern-.1667em\lower.7ex\hbox{E}\kern-.125emX}}
\begin{document}

\makeatletter
\newcommand{\linebreakand}{%
  \end{@IEEEauthorhalign}
  \hfill\mbox{}\par
  \mbox{}\hfill\begin{@IEEEauthorhalign}
}
\makeatother

\newcommand\correspondingauthor{\thanks{$^*$Corresponding author.}}

\title{ParvaGPU: Efficient Spatial GPU Sharing for Large-Scale DNN Inference in Cloud Environments}

\author[1]{Munkyu Lee}
\author[1]{Sihoon Seong}
\author[2]{Minki Kang}
\author[3]{Jihyuk Lee}
\author[4]{\\Gap-Joo Na}
\author[4]{In-Geol Chun}
\author[5]{Dimitrios Nikolopoulos}
\author[1]{Cheol-Ho Hong}

\affil[1]{Department of Intelligent Semiconductor Engineering, Chung-Ang University, Seoul, Republic of Korea}
\affil[2]{School of Computer Science and Engineering, Chung-Ang University, Seoul, Republic of Korea}
\affil[3]{School of Electrical and Electronics Engineering, Chung-Ang University, Seoul, Republic of Korea}
\affil[4]{Electronics and Telecommunications Research Institute, Daejeon, Republic of Korea}
\affil[5]{Department of Computer Science, Virginia Tech, Blacksburg, USA}

\maketitle

\begin{abstract}
In cloud environments, GPU-based deep neural network (DNN) inference servers are required to meet the Service Level Objective (SLO) latency for each workload under a specified request rate, while also minimizing GPU resource consumption. However, previous studies have not fully achieved this objective. In this paper, we propose ParvaGPU, a technology that facilitates spatial GPU sharing for large-scale DNN inference in cloud computing. ParvaGPU integrates NVIDIA's Multi-Instance GPU (MIG) and Multi-Process Service (MPS) technologies to enhance GPU utilization, with the goal of meeting the diverse SLOs of each workload and reducing overall GPU usage. Specifically, ParvaGPU addresses the challenges of minimizing underutilization within allocated GPU space partitions and external fragmentation in combined MIG and MPS environments. We conducted our assessment on multiple A100 GPUs, evaluating 11 diverse DNN workloads with varying SLOs. Our evaluation revealed no SLO violations and a significant reduction in GPU usage compared to state-of-the-art frameworks.
\end{abstract}


\section{Introduction} \label{introduction}
Rapid advancements in graphics processing units (GPUs) have significantly accelerated the spread of deep neural network (DNN)-based inference services~\cite{crankshaw2017clipper, olston2017tensorflow, gujarati2020serving, romero2021infaas} in areas such as real-time image analysis~\cite{ruan2020deep}, medical diagnostics~\cite{shen2017deep}, and natural language processing~\cite{li2018deep}. Major cloud companies, including Amazon and Google, are now offering instances equipped with high-speed GPUs~\cite{EC2_P4_instance}, enabling internet companies to deploy inference services flexibly without the need to directly purchase and manage GPU hardware~\cite{miao2023spotserve, wang2021morphling}. In a cloud environment, servers for DNN inference must satisfy the Service Level Objective (SLO) latency for each inference workload under a given request rate~\cite{Jones2016SRE, crankshaw2017clipper, gujarati2020serving, shen2019nexus}. Latency refers to the time taken to respond to a user's inference request, and request rate refers to the capacity to process inference requests per unit of time. While it is possible to allocate cloud resources generously to satisfy each workload's SLO, this approach leads to the issue of GPU resource wastage. For instance, although one could allocate one GPU per workload generously, the pay-per-use nature of cloud environments requires paying additional costs for any underutilized resources. Consequently, it is crucial to allocate only the minimum necessary GPU resources to each workload to meet its SLO, thereby maximizing GPU utilization and cost efficiency~\cite{ye2024deep}.

As a leader in the AI processor industry, NVIDIA provides Multi-Process Service (MPS)~\cite{nvidiaMPS} and Multi-Instance GPU (MIG)~\cite{nvidiaMIG} functionalities that enable the spatial partitioning of a single GPU for use by multiple workloads, thus maximizing the utilization of individual GPU resources~\cite{zhao2021survey}. MPS allows for the concurrent execution of GPU kernels generated by multiple CUDA processes on the GPU’s Streaming Multiprocessors (SMs), facilitating spatial sharing of the GPU among several processes. Since the introduction of the Volta architecture, MPS can proportionally divide spatial usage within a single GPU for each workload. However, because internal GPU resources such as caches and memory controllers are shared among workloads, this could lead to interference issues~\cite{chen2017prophet}. In contrast, the MIG feature can split one GPU into up to seven independent GPU instances, eliminating interference between workloads. However, each instance can only be configured with 1, 2, 3, 4, or 7 GPCs, composed of CUDA cores, tensor cores, and shared memory, limiting the ability to fine-tune spatial resource usage for each workload. Furthermore, with MIG enabled, a GPU can only be divided into 19 specific configurations, as shown in Figure~\ref{figure:MIG Configuration}, which presents constraints on flexibility when assigning various sizes of GPU instances~\cite{tan2021serving}.

When spatially partitioning GPUs to efficiently utilize them while ensuring the SLO for each workload, the following considerations are crucial~\cite{ye2024deep}. First, each workload must optimally utilize its allocated partition through high usage rates. Over-allocating GPU resources by setting a partition size larger than what is necessary for the respective workload leads to the wastage of GPU resources and increases the total number of GPUs utilized. This underutilization of the internal space of a partitioned GPU will henceforth be referred to as \emph{GPU internal slack} in this paper. Second, when deploying each GPU partition in a multi-GPU environment, it is essential to prevent fragmentation in the external space of the placed partitions. If the allocation process frequently results in non-continuous small spaces, precluding the assignment of larger-sized GPU partitions, this too will increase the total number of GPUs utilized. This phenomenon will be termed \emph{GPU external fragmentation}.

Previous research has yet to effectively resolve the challenge of ensuring the SLO for each workload while simultaneously reducing GPU resource consumption~\cite{dhakal2020gslice, choi2022serving,xu2022igniter,kim2022paris, tan2021serving}. gpulet~\cite{choi2022serving}, in an MPS environment, predicts performance degradation due to interference between workloads and adjusts the size of the GPU partition and batch size accordingly. However, gpulet's predictive model has limitations, allowing for the consolidation of only up to two workloads on a single GPU, which leads to significant internal slack by allocating all remaining GPU space to the second workload. iGniter~\cite{xu2022igniter} allocates additional GPU resources to each partition in an MPS environment to counteract the degradation in inference performance due to interference, achieved through interference modeling. Nevertheless, the developed model tends to allocate GPU resources generously to prevent SLO violations, resulting in internal slack, and does not account for external fragmentation, thus leading to GPU wastage. MIG-serving~\cite{tan2021serving}, in an MIG environment, considers the sizing and placement of GPU instances on actual GPUs as an NP-Hard problem and has developed the slow and fast algorithms for assigning instances across multiple GPUs. However, MIG-serving faces an issue of internal slack due to over-allocation of GPUs via heuristic algorithms. The degree to which each of these prior studies induces internal slack and external fragmentation will be discussed in the evaluation section.

In this paper, we propose ParvaGPU, an efficient GPU space-sharing technology that maximizes GPU utilization while supporting large-scale DNN inference in cloud environments, thereby enhancing cost efficiency. ParvaGPU combines MIG and MPS technologies to increase GPU utilization. ParvaGPU allocates partitioned MIG instances to each inference workload, preventing interference between different workloads. Within each MIG instance, ParvaGPU activates MPS and increases the number of processes for the same workload to maximize the utilization of resources within the MIG instance. In this paper, we refer to an MPS-activated MIG instance as \emph{GPU segment}. ParvaGPU develops the Segment Configurator to determine a set of GPU segments for each workload that meets the SLO while minimizing internal slack, based on profiled data; in cases of high request rates, multiple GPU segments are required as one segment may not suffice. Subsequently, the Segment Allocator distributes the combination of GPU segments for each model across multiple GPUs with minimal external fragmentation. Through an optimization process, even if external fragmentation occurs, it minimizes GPU usage by dividing and reallocating larger instances into smaller sizes.

To evaluate the performance of ParvaGPU, we utilized multiple Amazon p4de.24xlarge instances~\cite{EC2_P4_instance}, each equipped with eight A100 GPUs with 80GB of GPU memory. We conducted evaluations by varying the SLO latency and request rate across 11 different DNN workloads. The results showed that there were no violations of the SLO, and compared to state-of-the-art solutions, there was a substantial reduction in GPU usage across multiple scenarios. Although we demonstrate the effectiveness of spatial-sharing technology in ParvaGPU applied to DNN inference workloads, by modifying the SLO conditions in the developed algorithms, it can also be adapted for high-performance computing (HPC) applications and DNN training workloads.

To the best of our knowledge, ParvaGPU represents the first effort to highlight and simultaneously address the issues of internal slack and external fragmentation in an environment where MIG and MPS are blended. \cheolho{In this environment, finding the optimal combination of GPU segment configurations and their placement across multiple GPUs, which minimizes GPU usage while satisfying each workload’s SLO, is an NP-hard problem and highly challenging. To address this issue, ParvaGPU introduces algorithms that reduce the search space by dividing resource allocation and workload placement into two distinct stages.} The contributions of this paper are as follows:
\begin{itemize}
    \item In the Segment Configurator, we introduce the Optimal Triplet Decision algorithm, designed to prevent internal slack. This algorithm identifies GPU segments capable of delivering maximum throughput for each MIG instance size while considering the interference effects due to MPS among homogeneous workloads within the same MIG instance.
    \item We propose the Demand Matching algorithm within the Segment Configurator, optimized to rapidly derive the optimal set of GPU segments capable of satisfying high-volume request rates.
    \item In the Segment Allocator, we suggest the Segment Relocation algorithm, designed to allocate sets of GPU segments across multiple GPUs with minimal external fragmentation, while accommodating the complexities of MIG configurations (as illustrated in Figure~\ref{figure:MIG Configuration}).
    \item We offer the Allocation Optimization algorithm in the Segment Allocator, aimed at minimizing external fragmentation by splitting large segments into several smaller ones and reallocating them to empty spaces. This approach is especially effective in addressing external fragmentation that might persist even after executing the Relocation algorithm.
\end{itemize}

\begin{figure}[t]
    \centering
    \includegraphics[width=0.47\textwidth]{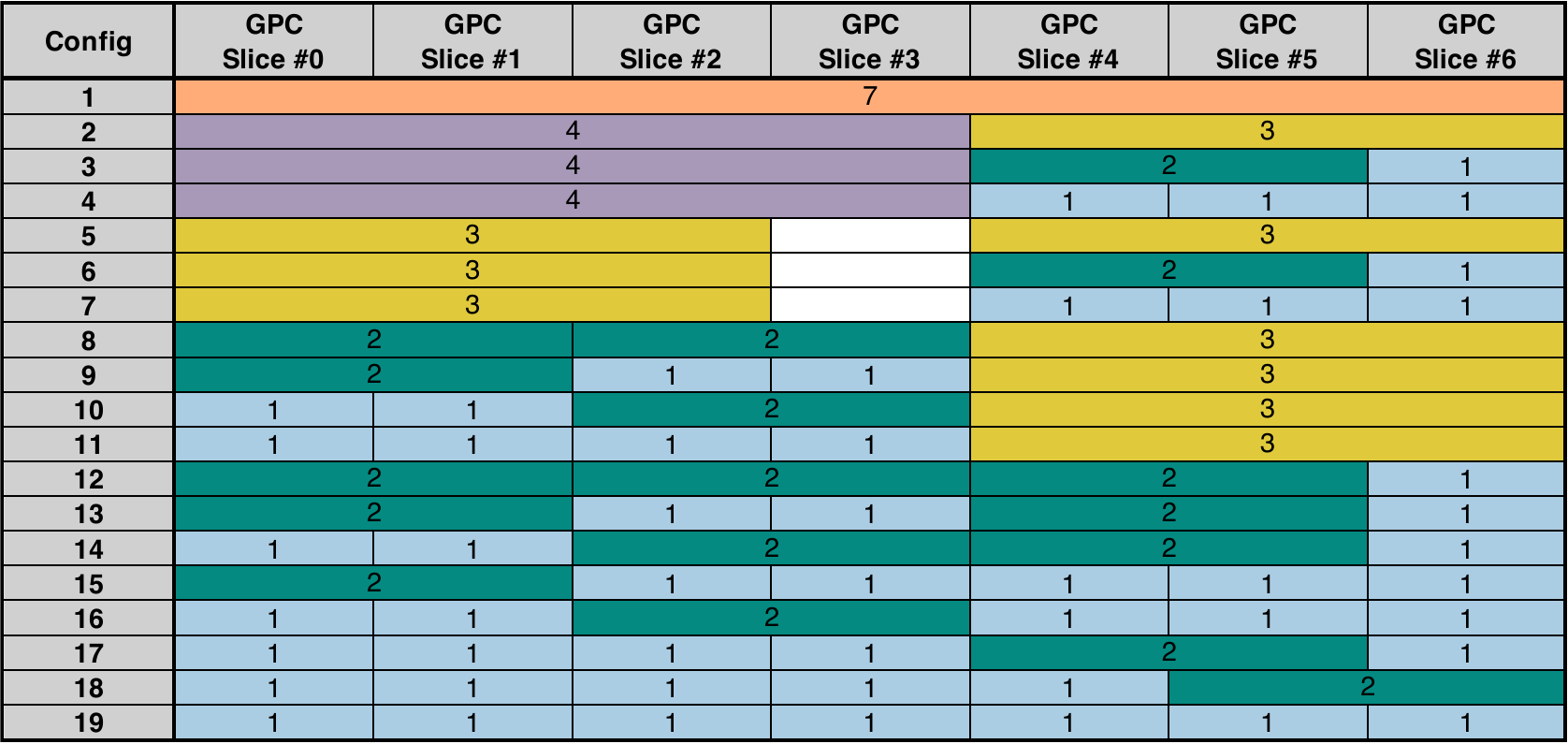}
    \caption{Supported MIG configurations on the NVIDIA A100 GPU.}
    \label{figure:MIG Configuration}
    \setlength{\belowcaptionskip}{-10pt}
\end{figure}

\section{Background and Related Work} \label{Background_and_Related_Work}

\begin{table*}[t]
\centering
\resizebox{\textwidth}{!}{
\begin{tabular}{cccccccc}
\hline
\textbf{} & \begin{tabular}[c]{@{}c@{}}MPS \\ support\end{tabular} & \begin{tabular}[c]{@{}c@{}}MIG \\ support\end{tabular} & \begin{tabular}[c]{@{}c@{}}Internal slack \\ prevention\end{tabular} & \begin{tabular}[c]{@{}c@{}}External fragmentation\\ prevention\end{tabular} & \begin{tabular}[c]{@{}c@{}}Spatial\\ scheduling\end{tabular} & \begin{tabular}[c]{@{}c@{}}High request\\ rate support\end{tabular} & \begin{tabular}[c]{@{}c@{}}Scheduling\\ overhead\end{tabular} \\ \hline
\begin{tabular}[c]{@{}c@{}}GSLICE~\cite{dhakal2020gslice}\end{tabular} & \ding{51} & \ding{55} & \ding{51} & \begin{tabular}[c]{@{}c@{}}\ding{55}\end{tabular} & \ding{51} & \ding{55} & Low \\
\begin{tabular}[c]{@{}c@{}}gpulet~\cite{choi2022serving}\end{tabular} & \ding{51} & \ding{55} & \ding{55} & N/A & 2 & \ding{51} & Medium \\
\begin{tabular}[c]{@{}c@{}}iGniter~\cite{xu2022igniter}\end{tabular} & \ding{51} & \ding{55} & \ding{55} & \ding{55} & \ding{51} & \ding{55} & Low \\
\begin{tabular}[c]{@{}c@{}}PARIS and ELSA~\cite{kim2022paris}\end{tabular} & \ding{55} & \ding{51} & \ding{55} & \ding{55} & N/A & \ding{55} & N/A \\
\begin{tabular}[c]{@{}c@{}}MIG-serving~\cite{tan2021serving}\end{tabular} & \ding{55} & \ding{51} & \ding{55} & \ding{51} & \ding{51} & \ding{51} & Very high \\
ParvaGPU & \ding{51} & \ding{51} & \ding{51} & \ding{51} & \ding{51} & \ding{51} & Low \\ \hline
\end{tabular}
}
\caption{Comparison of spatial GPU sharing solutions for inference servers.}
\label{table:Related Work}
\end{table*}

In this section, we examine NVIDIA's technologies for spatially sharing GPU resources as background. Furthermore, as related work, we analyze space-sharing frameworks for inference servers that utilize these technologies and compare these studies in Table~\ref{table:Related Work}.

\subsection{Multi-Process Service (MPS)}

Early NVIDIA GPUs allowed a single CUDA process to exclusively use GPU resources, leading to underutilization when the process's parallelism was insufficient. The MPS feature~\cite{nvidiaMPS} allows kernels from multiple processes to run concurrently on a single GPU by sharing space. Following the introduction of the Volta architecture, MPS has been enhanced to set spatial resource allocation quotas for each process. Although MPS partitions SMs for spatial sharing, internal GPU memory resources, including caches and memory controllers, are not isolated, resulting in interference between workloads~\cite{chen2017prophet}. The degree of this interference varies depending on the combination of workloads sharing the GPU, and its unpredictable performance can be a major cause of SLO violations. To address these issues, research such as GSLICE\cite{dhakal2020gslice}, gpulet~\cite{choi2022serving}, and iGniter~\cite{xu2022igniter} has been proposed.

GSLICE~\cite{dhakal2020gslice} adopts a self-tuning algorithm that adjusts the size of GPU partitions in MPS by measuring the latency and throughput of a workload to meet the SLO conditions. It also uses adaptive batching~\cite{wu2020irina} to determine a batch size that increases GPU utilization without violating the SLO. This approach allows GSLICE to prevent internal slack, but it does not address external fragmentation. Additionally, without considering multi-GPU environments, GSLICE is incapable of handling high request rates.

gpulet~\cite{choi2022serving} assigns two workloads to a GPU if the sum of their resource usage and additional resources considering interference does not exceed the GPU's total resource capacity. If this is not possible, a new GPU is allocated for processing. gpulet limits its allocation to a maximum of two workloads per GPU. The MPS partition is allocated based on the resource usage of the first workload, and the remaining GPU resources are then entirely assigned to the second workload's MPS partition. As such, while gpulet does not need to consider external fragmentation, this method can lead to overallocation in the second workload's partition, resulting in significant internal slack. Additionally, gpulet incurs a heavy overhead by performing profiling for all pairs of workloads to calculate interference.

iGniter~\cite{xu2022igniter} presents a model for predicting workload performance in an MPS environment, incorporating factors such as scheduling delays and L2 cache contention caused by interference. The coefficients of this model are determined through lightweight profiling. iGniter calculates the size of MPS partitions for each workload by adding the resource usage required to satisfy the workload's SLO to the additional resources needed for managing interference. However, to compensate for potential performance prediction errors arising from lightweight profiling, iGniter allocates additional GPU resources to each workload, leading to internal slack. Additionally, iGniter does not have a specific algorithm to address external fragmentation when assigning workloads to GPUs, and it lacks a mechanism for handling workloads with high request rates.

Compared to existing research in the MPS environment, ParvaGPU uniquely utilizes MPS for identical workloads within a single MIG instance. This methodology eliminates the requirement to forecast interference resulting from various heterogeneous workload combinations, consequently reducing the associated profiling overhead.

\subsection{Multi-Instance GPU (MIG)}

The MIG feature~\cite{nvidiaMIG}, introduced with NVIDIA's Ampere architecture GPUs, enables the division of a single GPU into up to seven independent GPU instances. Currently, the A30, A100, and H100 GPUs offer MIG functionality. Each MIG instance has separate L2 cache memory and memory controllers, offering predictable performance without interference from other workloads~\cite{choquette2021nvidia}. Instances can be reconfigured into independent GPU resources of 1, 2, 3, 4, or 7 GPCs. However, due to hardware limitations, configurations of 5 or 6 GPCs are not possible. The allocation of GPU memory for each instance can be divided according to predetermined configurations. For example, an NVIDIA A100 GPU with 80GB memory can have instances with 10, 20, 40, 40, 80GB of GPU memory, respectively. The possible instance configurations for a single GPU are limited, with only 19 combinations such as 1-1-1-1-1-1-1, 4-3, 4-2-1, and 4-1-1-1 being viable. These 19 configurations are depicted in Figure~\ref{figure:MIG Configuration}. Determining the size of MIG instances to meet the SLO for each workload and placing these combinations of instances across multiple GPUs presents a challenge~\cite{li2022miso, li2023clover}. To address this, research including PARIS and ELSA~\cite{kim2022paris}, and MIG-serving~\cite{tan2021serving} has been proposed.

PARIS and ELSA~\cite{kim2022paris} feature a dual approach where PARIS determines suitable MIG instance sizes for each workload based on the batch size's normal distribution, and ELSA schedules workloads temporally on GPUs that have been heterogeneously partitioned into MIG instances. PARIS informs about the appropriate MIG Instance size for each workload but does not include a mechanism to prevent external fragmentation during GPU allocation. ELSA focuses on optimizing temporal utilization rather than spatial scheduling within a single MIG partition, thus not addressing the issue of internal slack.

MIG-serving~\cite{tan2021serving} views the entire process of finding suitable MIG instance sizes for workloads and allocating these instances to GPUs as a cutting stock problem~\cite{fang2023solving}, which is an NP-hard problem. MIG-serving's Optimizer tackles this challenge using various algorithms: a greedy algorithm (fast algorithm), a genetic algorithm (slow algorithm), and a Monte Carlo tree search algorithm (slow algorithm). However, the Optimizer can lead to internal slack by over-allocating GPU resources to workloads based on heuristic scores during initial stages. Moreover, the simultaneous execution of finding appropriate instance sizes and allocating instances on multiple GPUs, even with the fast algorithm, results in significant scheduling overhead as the number of models increases.

ParvaGPU significantly reduces scheduling overhead by partitioning the process of utilizing MIG into two separate stages: the GPU Segment Configurator and Allocator. Both stages are computationally lightweight.

\section{Design} \label{Design}
In this section, we describe the overall architecture of ParvaGPU, then examine the characteristics of DNN inference workloads. Following this, we explain the key components, the GPU Segment Configurator and the GPU Segment Allocator, and discuss the deployment method. Finally, we analyze the time complexity of each module.

\subsection{Overall Design}

\begin{figure}[t]
    \centering
    \includegraphics[width=0.47\textwidth]{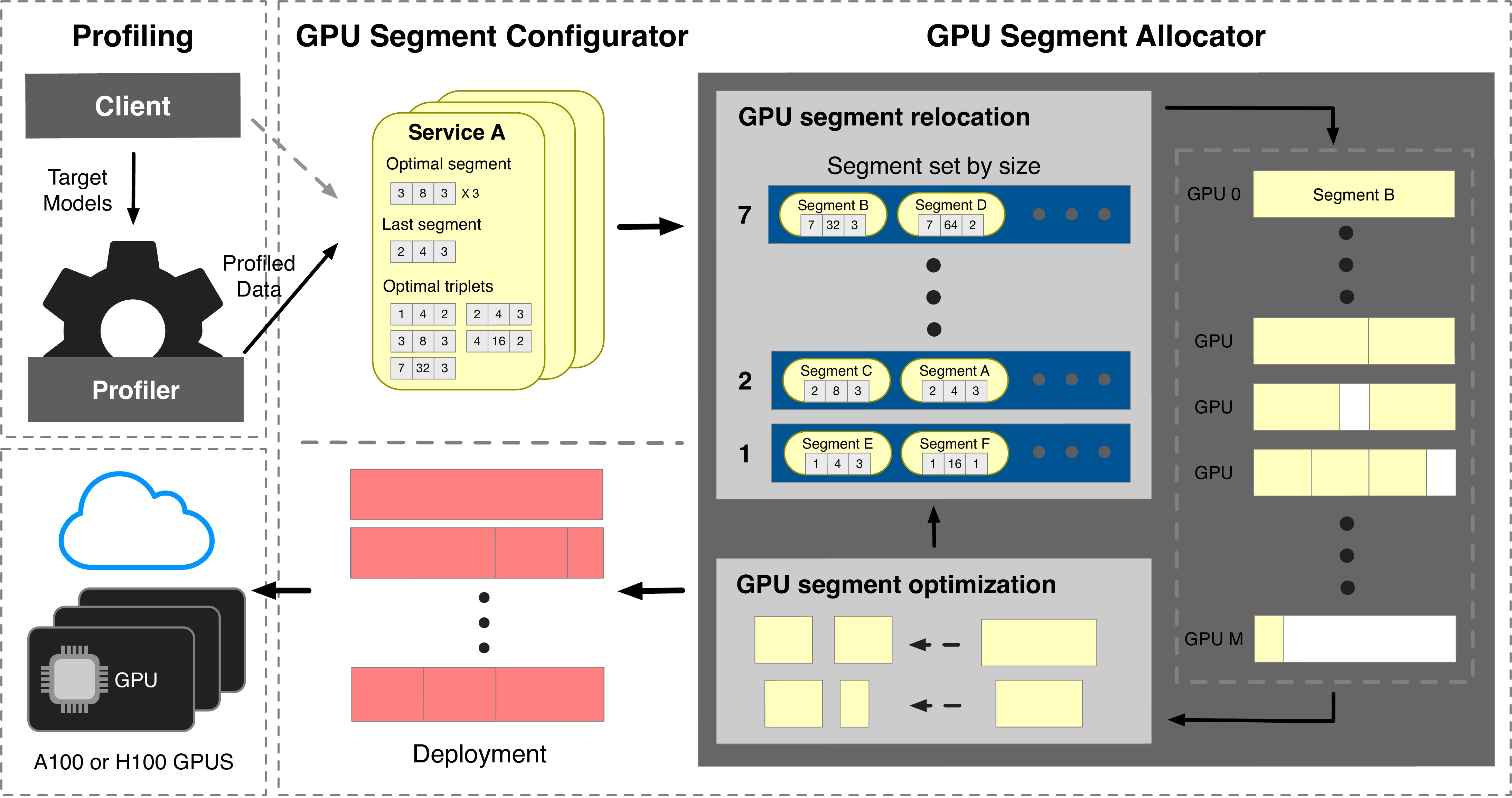}
    \caption{Overall design of ParvaGPU.}
    \label{fig:overall design}
\end{figure}

ParvaGPU provides an optimized spatial GPU sharing technology for inference workloads. It combines NVIDIA's MIG and MPS technologies to finely partition GPU space. ParvaGPU assigns divided MIG instances to each inference workload, thereby preventing mutual interference between different workloads. It activates MPS in each MIG instance and increases the number of processes for the same workload to maximize resource utilization within the MIG instance. We refer to an MPS-enabled MIG instance as a \emph{GPU segment}. To use the minimum amount of GPU while satisfying the SLO for each workload, it is crucial to minimize both GPU internal slack and GPU external fragmentation within the bounds of meeting each workload's SLO.

Figure~\ref{fig:overall design} presents the entire architecture of ParvaGPU. A client provides DNN models for the service, inclusive of their SLOs. The Profiler then proceeds to evaluate throughput and latency for various MIG instance sizes, model batch sizes, and the number of processes in MPS. Based on this profiling data, the GPU Segment Configurator determines the optimal set of GPU segments that align with the SLOs of the workloads. Subsequently, the GPU Segment Allocator allocates these predetermined GPU segments to each GPU, aiming to minimize external fragmentation. The final deployment, as orchestrated by ParvaGPU, enables servicing of all provided workloads in a cloud environment with maximized GPU utilization, without any SLO violations.

\begin{figure}[t]
    \centering
    \includegraphics[width=0.47\textwidth]{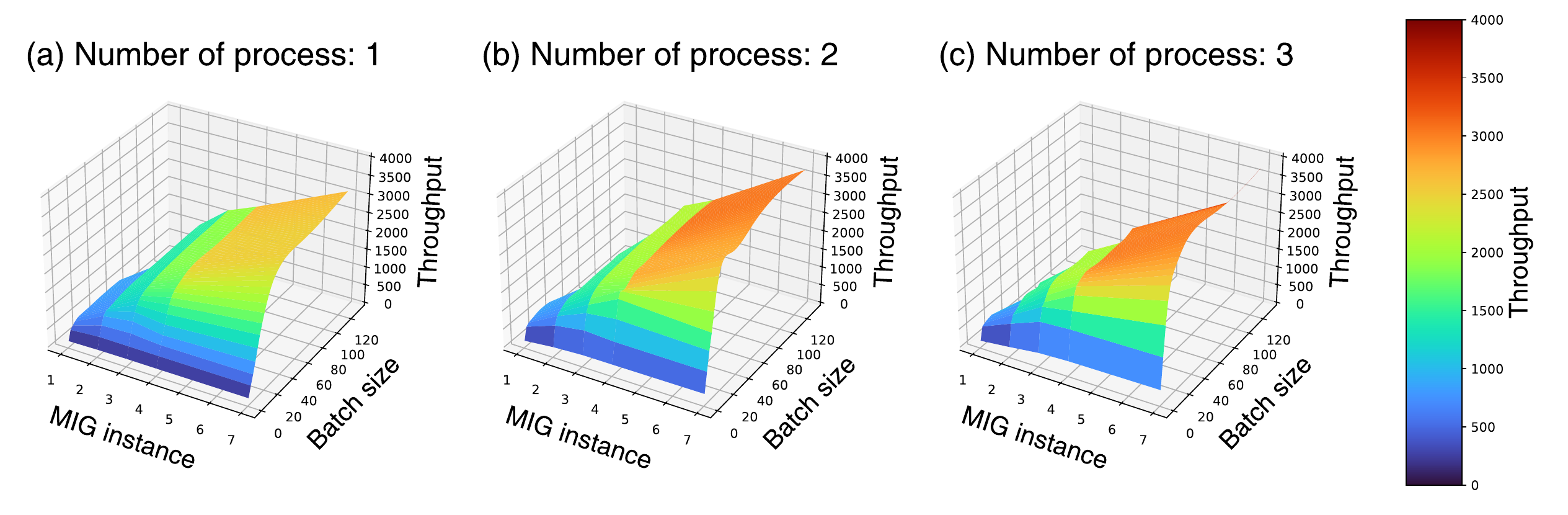}
    \caption{Throughput (requests/s) of InceptionV3 with different batch sizes and instance sizes for each process count of 1 (a), 2 (b), and 3 (c).}
    \label{fig:InceptionV3 Throughput}
    \includegraphics[width=0.47\textwidth]{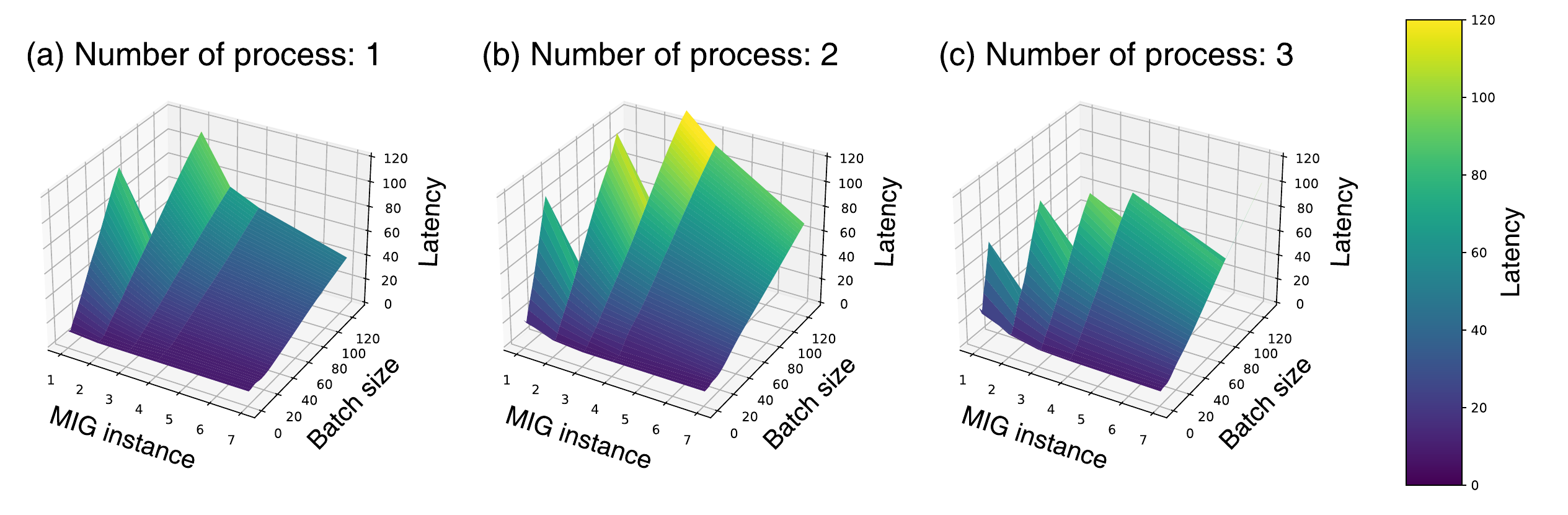}
    \caption{Latency (ms) of InceptionV3 with different batch sizes and instance sizes for each process count.}
    \label{fig:InceptionV3 Latency}
\end{figure}

\subsection{Workload Characteristic Analysis}
In this chapter, we explore how the size of the MIG instance and the number of MPS processes, in conjunction with the model batch size, impact the performance of a workload, using InceptionV3~\cite{xia2017inception} as an illustrative example. Other DNN models were observed to exhibit similar characteristics.

Figure~\ref{fig:InceptionV3 Throughput} shows the throughput of InceptionV3 as assessed by varying the MIG instance sizes and model batch sizes under different quantities of MPS processes: specifically, 1 process in scenario (a), 2 processes in scenario (b) and 3 processes in scenario (c). Although MIG instance sizes 5 and 6 do not exist, interpolation has been used to ensure smoother transitions in the graph. Points where the GPU memory allocated to each instance size was insufficient, leading to out-of-memory errors, are not included in the graph. In general, increases in MIG instance size, model batch size, and number of MPS processes correlate with improvements in throughput. A key observation from this experiment is that with a fixed MIG instance size, larger batch sizes can lead to diminishing returns in performance gains as the number of processes increases, and vice versa. Thus, it is crucial to find the optimal interaction point between these two factors, where their combined increase results in a balanced impact on performance.

Figure~\ref{fig:InceptionV3 Latency} displays the latency changes for the same scenarios depicted in Figure~\ref{fig:InceptionV3 Throughput}. It is observed that latency decreases as the instance size increases, while increasing both the batch size and the number of processes leads to higher latency. A significant finding from this experiment is that when an MIG instance allocated to workloads is already highly utilized, efforts to increase throughput by increasing batch size or process count result in a disproportionately large increase in latency. For instance, when the instance size is set to 1 and the batch size to 4, increasing the number of processes from 1 to 2 and 3 results in a slight improvement in throughput, with figures of 354, 444, and 446, respectively. However, compared to this, the latency increases significantly: from 11ms with one process to 18ms with two processes, and 27ms with three processes, representing increases of 1.6 and 2.45 times, respectively. In contrast, with an instance size of 4 and a batch size of 8, increasing the number of processes leads to a significant increase in throughput (786, 1695, and 1810, respectively), but the corresponding increases in latency are minimal, with values of 10ms, 9ms, and 13ms, respectively. Therefore, to improve throughput, it is advantageous to increase the number of instances, ensuring that there are no violations of SLO latency.

\subsection{Profiler}
\label{Profiler}
Users only need to perform profiling once for a short duration when initially registering a service with ParvaGPU. ParvaGPU's Profiler, utilizing either a single GPU or multiple GPUs, records the throughput and latency of each workload by varying the instance size, batch size, and number of processes for the workload. The instance size is limited to five options (1, 2, 3, 4, and 7 GPCs), but the batch size and the number of processes increase incrementally, with no defined upper limit. To avoid exhaustive search, we suggest using a set of eight common batch sizes, exponentially increasing from 1 to 128, and limit the number of processes to three, considering out-of-memory scenarios within the MIG instance. Previous studies also conduct profiling for each model~\cite{dhakal2020gslice, choi2022serving,xu2022igniter,kim2022paris, tan2021serving}. iGniter~\cite{xu2022igniter} employs sampling-based lightweight profiling but faces accuracy limitations. gpulet~\cite{choi2022serving} profiles workload pairs in an MPS environment, leading to considerable overhead. In contrast, ParvaGPU performs profiling only at essential points and does not require profiling of different workload pairs in an MPS environment. Consequently, this approach significantly reduces the time required for profiling while ensuring the accuracy of the profiling results.

\subsection{GPU Segment Configurator}
\label{GPU Segment Configurator}

The primary goal of the GPU Segment Configurator in ParvaGPU is to identify a set of GPU segments that not only meets each workload's SLO but also minimizes internal slack. To address the issue of an increase in search space resulting from considering both of these elements together, ParvaGPU employs a two-step algorithmic approach: the Optimal Triplet Decision and the Demand Matching algorithms.

\subsubsection{Optimal Triplet Decision}
\label{Optimal Triplet Decision}

\begin{table}[t]
\centering
\resizebox{\columnwidth}{!}{%
\begin{tabular}{p{1.8cm} p{4cm}}
\hline
Variable name & Description \\ 
\hline
$id$ & Service identification number \\
$lat$ & SLO latency \\
$req\_rate$ & Request rate\\
$opt\_tri\_array$ & Optimal triplet array \\
$opt\_seg$ & Optimal segment \\
$num\_opt\_seg$ & Number of optimal segments \\
$last\_seg$ & Last segment \\
\hline
\end{tabular}%
}
\caption{Member variables for the service object.}
\label{tab:service object}
\end{table}

\begin{algorithm}[t]
\DontPrintSemicolon
\SetAlgoLined
\SetNlSty{textbf}{}{:}
\SetAlgoNlRelativeSize{-1}
\SetKwInOut{Input}{Input}
\SetKwInOut{Output}{Output}
\SetKw{KwBy}{by}
\SetKwProg{Fn}{Function}{}{}
\label{algorithm:configurator}

\caption{GPU Segment Configurator algorithm}
/* $S$ and $P$ are object arrays, where $S$ represents the set of services, and $P$ represents the profile results. The term \emph{lat} stands for latency, and \emph{tp} denotes throughput. */\;
\Fn{\textsc{TripletDecision}}{
    \Input{$S$, $P$}
    \Output{$S$} 
    \For{$i=1$ \textbf{to} sizeof($S$)}{
        $max\_triplets \gets \text{empty array}$ \;
        \For{$j=1$ \textbf{to} sizeof($P$)}{
            \If{$S[i].lat > P[j].lat$}{
                \textsc{UpdateMaxTriplets($max\_triplets, P[j]$)} 
            }
        }
        $S[i].opt\_tri\_array \gets max\_triplets$  \;
    }
    \Return $S$
}
\;
\Fn{\textsc{DemandMatching}}{
    \Input{$S$}
    \Output{$S$}
    \For{$i=1$ \textbf{to} sizeof($S$)}{
        $left\_req\_rate \gets 0$ \;
        $S[i].opt\_seg \gets \textsc{OptSeg($S[i].opt\_tri\_array$)}$\;
        $S[i].num\_opt\_seg \gets \lfloor \frac{S[i].req\_rate}{S[i].opt\_seg.tp} \rfloor$ \;
        $left\_req\_rate \gets \textsc{GetLeftReqRate($S[i]$)}$ \;
        $S[i].last\_seg \gets \textsc{LastSeg($left\_req\_rate, S[i].opt\_tri\_array$)}$\;
        
    }
    \Return $S$
}
\end{algorithm}

The Optimal Triplet Decision algorithm identifies the points of maximum throughput for each of five instance sizes, thereby deriving a total of five optimal triplets. Each triplet consists of an instance size, a batch size, and a process size. ParvaGPU allows for the configuration of a single service using various sizes of GPU segments in order to prevent internal slack and external fragmentation. For this purpose, it determines the optimal triplets for each of the instance sizes. 

Algorithm~\ref{algorithm:configurator} describes the Segment Configurator, which includes the Optimal Triplet Decision algorithm. Table~\ref{tab:service object} displays the member variables of the service object used in the algorithms of this paper. Each service object, a part of the complete service set $S$, encompasses an identification number, SLO latency, request rate of the service, and variables for storing the outcomes of the Segment Configurator's execution. The \textsc{TripletDecision} function performs a comprehensive search for each service based on the input profiling results and the entire set of services $S$ (lines 3-12). This function only activates the \textsc{UpdateMaxTriplets} function for profiling results that demonstrate latencies lower than the service's SLO latency (line 6). The \textsc{UpdateMaxTriplets} function identifies the batch size and process size that achieve maximum throughput for each of the five instance sizes, and updates this information in the $max\_triplets$ array. This array is then saved in the service object's optimal triplet array (line 10).

\subsubsection{Demand Matching}
The Demand Matching algorithm is designed to accommodate high request rates that are difficult to handle with a single GPU segment, while using the minimum necessary GPU resources. This is achieved by combining the optimal triplet array, derived from a prior stage, to determine the best set of GPU segments for each service. We define this challenge as a type of tree search problem~\cite{cen2022tree}. In this tree, a node represents the remaining request rate, which is the total request rate minus the throughput achieved with the resources allocated up to that point. An edge consists of five throughput options based on the optimal triplets that can be selected at the current node. The root node of the tree represents the total request rate that the service needs to fulfill, and the tree expansion stops at a node when the remaining request rate becomes zero or less during the tree search. The goal of this problem is to find the path from the root node to a leaf node that achieves this at the minimum cost.

The Demand Matching algorithm strives to find an efficient path that utilizes the fewest possible number of GPCs, without necessitating a full traversal of the tree's nodes. During the tree search, the most efficient path is determined by selecting edges that maximize throughput per instance size (or the number of GPCs). This is proven in the following demonstration.

\begin{equation}\label{eqn:num_inst}
\text{Number of Segments} = \frac{Request\ rate}{\sum_{l=1}^{H} (Throughput_l)}
\end{equation}

Equation~\ref{eqn:num_inst} calculates the number of segments required to meet the request rate of a service by expanding the tree from level 1 to H, where H is the height of the tree and considering the throughput of the selected edges at each level.

\begin{equation}\label{eqn:num_gpc}
\text{Number of GPCs} = {Request\ rate} \times \sum_{l=1}^{H} (\frac{Instance\ size_l}{Throughput_l})
\end{equation}

Reflecting the variation in the number of GPCs used per instance size, Equation~\ref{eqn:num_gpc} multiplies both sides of Equation~\ref{eqn:num_inst} by the instance size to derive the required number of GPCs. Given that $Request\ rate$ is a constant value, minimizing $Number\ of\ GPCs$ requires selecting an instance size where the ratio $Throughput/Instance\ size$ is at its maximum. The Demand Matching algorithm determines the optimal segment by identifying the triplet where this value is maximized. This approach results in an excellent time complexity of $O(1)$.

In Algorithm~\ref{algorithm:configurator}, the \textsc{DemandMatching} function utilizes the \textsc{OptSeg} function to find the triplet that maximizes $Throughput/Instance\ size$, and assigns it to the optimal segment variable of the service object (line 17). The request rate is then divided by the throughput of the identified optimal segment, and the floor function is applied to determine the necessary number of optimal segments, excluding the edge leading to the last tree node (line 18). The last segment is chosen to be the smallest instance size that can fulfill the remaining request rate (lines 19-20). \cheolho{The \textsc{GetLeftReqRate} function returns the remaining request rate for the last tree node, while the \textsc{LastSeg} function identifies the smallest instance size that can satisfy the remaining request rate.} When the final segment is also filled with an optimal segment, internal slack may occur if the processing capacity required by the leaf node is low. The Demand Matching algorithm is also efficient for small request rates that can be handled by a single segment. In such cases, the floor function in line 18 returns the number of optimal segments as zero, and lines 19-20 enable the selection of a segment suitable for that particular request rate.

\subsection{GPU Segment Allocator}
\label{GPU Segment Allocator}
The objective of the GPU Segment Allocator is to create a deployment map that positions the collective segments of all services across multiple GPUs while minimizing external fragmentation. The actual allocation to the physical GPUs is performed after the execution of the GPU Segment Allocator is complete. Similar to the Segment Configurator, ParvaGPU adopts a two-stage algorithmic strategy to reduce the search space: the Segment Relocation and Allocation Optimization algorithms. 

\subsubsection{Segment Relocation}

\begin{table}[t]
\centering
\resizebox{\columnwidth}{!}{%
\begin{tabular}{p{1.8cm} p{4cm}}
\hline
Variable name & Description \\ 
\hline
$id$ & GPU identification number \\
$num\_gpcs$ & Number of allocated GPCs\\
$seg\_array$ & Array of allocated segment objects \\
\hline
\end{tabular}%
}
\caption{Member variables for the GPU object.}
\label{tab:gpu object}
\end{table}

\begin{algorithm}[t]
\DontPrintSemicolon
\SetAlgoLined
\SetNlSty{textbf}{}{:}
\SetAlgoNlRelativeSize{-1}
\SetKwInOut{Input}{Input}
\SetKwInOut{Output}{Output}
\SetKw{KwBy}{by}
\SetKwProg{Fn}{Function}{}{}
\label{algorithm:allocator}

\caption{GPU Segment Allocator algorithm}
/* $G$ is object arrays representing the allocated GPUs with mapped segments, and \emph{tp} denotes throughput. */\;

\Fn{\textsc{SegmentRelocation}}{
    \Input{$S$}
    \Output{$G$}
    
    \For{$i=1$ \textbf{to} sizeof($S$)}{ 
        \For{$j=1$ \textbf{to} $S[i].num\_opt\_seg$}{
            \textsc{Enqueue($S[i].id, S[i].opt\_seg$)}\;
        }
        \textsc{Enqueue($S[i].id, S[i].last\_seg$)}\;
    }
    $G \gets $\textsc{Allocation()}\;
    \Return $G$\;
}
\;
\Fn{\textsc{AllocationOptimization}}{
    \Input{$G$}
    \Output{Optimized $G$}

    $freed\_rate \gets \text{empty array}$ \;
    \For{$i=sizeof({G})$ \textbf{to} $1$}{
        \If{$G[i].num\_gpcs \leq 4$}{
            \For{$j=1$ \textbf{to} sizeof($G[i].seg\_array)$}{
                $small\_segs \gets \ \text{empty array}$ \;
                $s \gets G[i].seg\_array[j].service$\;
                $tp \gets G[i].seg\_array[j].tp$\;
                
                $freed\_rate[s.id] \mathrel{{+}{=}} tp$\;
                \textsc{FreeSegment($G[i].seg\_array[j]$)}\;
                
                $small\_segs \gets $  \textsc{SmallSegments($s.id, freed\_rate[s.id]$)}\;
                
                \For{$k=1$ \textbf{to} $sizeof(small\_segs)$}{
                    $freed\_rate[s.id] \mathrel{{-}{=}} small\_segs[k].tp$\;
                    \textsc{Enqueue($s.id, small\_segs[k]$)}\; 
                }
            }
        }
        $G = $\textsc{Allocation()}\;
    }
    \Return $G$
}
\end{algorithm}

The Segment Relocation algorithm creates a deployment map that allocates the segments of various services to multiple GPUs, reflecting the complexities of the MIG configurations and aiming to minimize external fragmentation as much as possible. To rapidly address this issue without exhaustive search, we employed a heuristic approach: sorting the segments of all services by size and then allocating them to GPUs in descending order. This technique is frequently used to solve problems similar to the irregular object-packing problem~~\cite{leao2020irregular}.

Algorithm~\ref{algorithm:allocator} elucidates the GPU Segment Allocator, inclusive of the Segment Relocation algorithm. The \textsc{SegmentRelocation} function places the optimal segments of each service, in their respective quantities, as well as the last segment, into queues organized by instance size (lines 3-8). \cheolho{The \textsc{Enqueue} function analyzes the size of the segment passed as an argument and places it in the corresponding queue based on its size.} Subsequently, the \textsc{Allocation} function determines the allocation of the segments from each queue to GPUs (line 9). This function processes the queues starting with those containing larger segment sizes and moves to smaller queues as each becomes empty. Upon extracting a segment from a queue, the function begins by searching from the first GPU, assessing whether the current GPU can accommodate that segment. Even if there is space for the segment in the current GPU, the decision is made to place it in that GPU or in the next available GPU, taking into account the constraints of the MIG configurations.

In the \textsc{Allocation} function for GPU allocation, constraints due to MIG configurations are as follows. Assuming that A100 and H100 GPUs can accommodate a total of 7 MIG instances, let us consider 7 slots available (numbered 0-6), as shown in Figure~\ref{figure:MIG Configuration}. We then assess where each segment size, decreasing from 7 to 1, can be placed. Segments of sizes 7 and 4 are limited to position only in slot 0. Size 3 segments could be allocated to either slot 0 or slot 4, but positioning these segments in slot 0 is generally not advisable. Placing a size 3 segment in slot 0 prevents the allocation of a size 1 segment in slot 3 due to the constraints of configurations 5 through 7 shown in Figure~\ref{figure:MIG Configuration}, which can cause significant external fragmentation across multiple GPUs. Therefore, priority is given to allocating size 3 segments in slot 4. Size 2 segments can be placed in slots 0, 2, 4, or 5, but given the higher demand for size 3 segments, size 2 segments are preferably allocated to slots 0 or 2, avoiding slots 4 and 5. Finally, size 1 segments are initially placed in slots 0-3 and then 4-6 to avoid interfering with the allocation of size 3 segments.

\subsubsection{Allocation Optimization}
The Allocation Optimization algorithm comprehensively minimizes external fragmentation resulting from small empty spaces left after segment placement by the Segment Relocation algorithm. To resolve this, the algorithm begins with the last GPU, targeting those GPUs with a higher degree of fragmentation. It divides the larger segments of these GPUs into smaller segments and then reallocates these smaller segments to the empty spaces, starting from the front GPUs.

In Algorithm~\ref{algorithm:allocator}, the \textsc{AllocationOptimization} function receives a set of GPU objects, denoted as $G$, which is the return value of the \textsc{SegmentRelocation} function. The attributes of the GPU object are detailed in Table~\ref{tab:gpu object}, including the GPU identification number, the count of allocated GPCs, and an array of allocated segment objects. The \textsc{AllocationOptimization} function starts with the last GPU, dividing the segments of those GPUs where the total count of allocated GPCs falls below a specified threshold (lines 13-15). In such cases, it is considered that the GPU has a high level of fragmentation. This threshold value is adjustable depending on the environment; in this paper, it is heuristically set to 4 for optimal fragmentation minimization. The function iterates through the segments of the targeted GPU, sums the throughput of the segments to be released into the $freed\_rate$ array indexed by service id, and then releases these segments \cheolho{using the \textsc{FreeSegment} function.} (lines 16-21). Next, the \textsc{SmallSegments} function retrieves small segments of size 1 or 2 that can satisfy the released throughput, based on the optimal triplet information of the service object, and stores them in the $small\_segs$ array (line 22). The total throughput handled by these small segments is then subtracted from $freed\_rate$, and the small segments are queued for allocation (lines 23-26). As the total throughput of $small\_segs$ usually exceeds that of the released segments, the surplus is reflected in the \textsc{SmallSegments} function for the next GPU to ensure the allocation of the smallest number of segments possible. Once the processing of a single GPU is complete, the \textsc{Allocation} function is called. This involves reallocating the $small\_segs$ array and updating the fragmentation level of each GPU (line 29). The final result returned by this function is Optimized $G$, a deployment map with minimal external fragmentation.

\subsection{Deployment}

ParvaGPU, upon receiving an optimized deployment map, optimized $G$, from the Segment Allocator, reconfigures the MIG and MPS of the physical GPUs and then launches inference servers to provide services. ParvaGPU can flexibly respond to changes in the SLO of a service. Re-profiling of the model is unnecessary; the Segment Configurator reconstructs only the optimal segments and the last segment for the service. This service is then removed from the deployment map $G$, and a segment relocation process is specifically carried out for it to acquire a new $G$. Following this, a segment optimization process leads to an optimized $G$ with minimized external fragmentation. This method minimizes the overhead of reconfiguration, as services whose placement has not changed do not require reconfiguration. To prevent service disruptions during brief periods of reconfiguration of MIG and MPS, which can range from milliseconds to a few seconds, services undergoing reconfiguration can continue operating using shadow processes on spare GPUs. We plan to explore the topic of reducing reconfiguration overhead using shadow processes in our subsequent research.

\subsection{Time Complexity}
ParvaGPU efficiently reduces the search space by dividing resource allocation and workload placement into two stages. Specifically, the Segment Configurator Algorithm~\ref{algorithm:configurator} has a time complexity of $O(NIBP)$ (where N is the number of services, I is the number of instance sizes, B is the number of batch sizes, and P is the number of processes). As described in Section~\ref{Profiler}, $I$ is 5, $B$ is 8, and $P$ is 3, which simplifies the complexity of this stage to $O(N)$. The Segment Allocator Algorithm~\ref{algorithm:allocator} has a time complexity of $O(NS) + O(NMK)$ (where S is the number of optimal triplets per service, M is the number of assigned GPUs, and K is the number of segments per GPU). Considering that a maximum of seven segments can be placed on a single GPU, $K$ can be treated as a constant, making the complexity of this stage $O(NS) + O(NM)$.

\section{Evaluation} \label{Evaluation}
\begin{table*}[t]
\centering

\resizebox{\textwidth}{!}{%
\begin{tabular}{cc|ccccccccccc}
\hline
 \multicolumn{2}{c|}{Workload features}  & BERT-large & DenseNet-121 & DenseNet-169 & DenseNet-201 & InceptionV3 & MobileNetV2 & ResNet-101 & ResNet-152 & ResNet-50 & VGG-16 & VGG-19 \\ \hline
 \multicolumn{2}{c|}{Number of parameters} & 330M & 8.0M & 14.1M & 20.0M & 27.2M & 3.5M & 44.5M & 60.2M & 25.6M & 138.4M & 143.7M \\ \hline
Scenario 1 (S1) &
  \begin{tabular}[c]{@{}c@{}}Request rate\\ Latency\end{tabular} &
  \begin{tabular}[c]{@{}c@{}}19\\ 6,434\end{tabular} &
  \begin{tabular}[c]{@{}c@{}}353\\ 183\end{tabular} &
  N/A &
  N/A &
  \begin{tabular}[c]{@{}c@{}}460\\ 419\end{tabular} &
  \begin{tabular}[c]{@{}c@{}}677\\ 167\end{tabular} &
  N/A &
  N/A &
  \begin{tabular}[c]{@{}c@{}}829\\ 205\end{tabular} &
  N/A &
  \begin{tabular}[c]{@{}c@{}}354\\ 397\end{tabular} \\ \hline
Scenario 2 (S2) &
  \begin{tabular}[c]{@{}c@{}}Request rate\\ Latency\end{tabular} &
  \begin{tabular}[c]{@{}c@{}}19\\ 6,434\end{tabular} &
  \begin{tabular}[c]{@{}c@{}}353\\ 183\end{tabular} &
  \begin{tabular}[c]{@{}c@{}}308\\ 217\end{tabular} &
  \begin{tabular}[c]{@{}c@{}}276\\ 169\end{tabular} &
  \begin{tabular}[c]{@{}c@{}}460\\ 419\end{tabular} &
  \begin{tabular}[c]{@{}c@{}}677\\ 167\end{tabular} &
  \begin{tabular}[c]{@{}c@{}}393\\ 212\end{tabular} &
  \begin{tabular}[c]{@{}c@{}}281\\ 213\end{tabular} &
  \begin{tabular}[c]{@{}c@{}}829\\ 205\end{tabular} &
  \begin{tabular}[c]{@{}c@{}}410\\ 400\end{tabular} &
  \begin{tabular}[c]{@{}c@{}}354\\ 397\end{tabular} \\ \hline
Scenario 3 (S3) &
  \begin{tabular}[c]{@{}c@{}}Request rate\\ Latency\end{tabular} &
  \begin{tabular}[c]{@{}c@{}}46\\ 4,294\end{tabular} &
  \begin{tabular}[c]{@{}c@{}}728\\ 126\end{tabular} &
  \begin{tabular}[c]{@{}c@{}}633\\ 150\end{tabular} &
  \begin{tabular}[c]{@{}c@{}}493\\ 119\end{tabular} &
  \begin{tabular}[c]{@{}c@{}}1,051\\ 282\end{tabular} &
  \begin{tabular}[c]{@{}c@{}}1,546\\ 113\end{tabular} &
  \begin{tabular}[c]{@{}c@{}}760\\ 144\end{tabular} &
  \begin{tabular}[c]{@{}c@{}}543\\ 146\end{tabular} &
  \begin{tabular}[c]{@{}c@{}}1,463\\ 138\end{tabular} &
  \begin{tabular}[c]{@{}c@{}}780\\ 227\end{tabular} &
  \begin{tabular}[c]{@{}c@{}}673\\ 265\end{tabular} \\ \hline
Scenario 4 (S4) &
  \begin{tabular}[c]{@{}c@{}}Request rate\\ Latency\end{tabular} &
  \begin{tabular}[c]{@{}c@{}}69\\ 4,294\end{tabular} &
  \begin{tabular}[c]{@{}c@{}}1,091\\ 126\end{tabular} &
  \begin{tabular}[c]{@{}c@{}}949\\ 150\end{tabular} &
  \begin{tabular}[c]{@{}c@{}}739\\ 119\end{tabular} &
  \begin{tabular}[c]{@{}c@{}}1,576\\ 282\end{tabular} &
  \begin{tabular}[c]{@{}c@{}}2,318\\ 113\end{tabular} &
  \begin{tabular}[c]{@{}c@{}}1,140\\ 144\end{tabular} &
  \begin{tabular}[c]{@{}c@{}}815\\ 146\end{tabular} &
  \begin{tabular}[c]{@{}c@{}}2,195\\ 138\end{tabular} &
  \begin{tabular}[c]{@{}c@{}}1,169\\ 227\end{tabular} &
  \begin{tabular}[c]{@{}c@{}}1,010\\ 265\end{tabular} \\ \hline
Scenario 5 (S5) &
  \begin{tabular}[c]{@{}c@{}}Request rate\\ Latency\end{tabular} &
  \begin{tabular}[c]{@{}c@{}}843\\ 2,153\end{tabular} &
  \begin{tabular}[c]{@{}c@{}}2,228\\ 69\end{tabular} &
  \begin{tabular}[c]{@{}c@{}}3,507\\ 84\end{tabular} &
  \begin{tabular}[c]{@{}c@{}}1,513\\ 70\end{tabular} &
  \begin{tabular}[c]{@{}c@{}}3,815\\ 146\end{tabular} &
  \begin{tabular}[c]{@{}c@{}}5,009\\ 59\end{tabular} &
  \begin{tabular}[c]{@{}c@{}}1,874\\ 77\end{tabular} &
  \begin{tabular}[c]{@{}c@{}}1,340\\ 80\end{tabular} &
  \begin{tabular}[c]{@{}c@{}}2,796\\ 72\end{tabular} &
  \begin{tabular}[c]{@{}c@{}}1,773\\ 115\end{tabular} &
  \begin{tabular}[c]{@{}c@{}}1,531\\ 134\end{tabular} \\ \hline
Scenario 6 (S6) &
  \begin{tabular}[c]{@{}c@{}}Request rate\\ Latency\end{tabular} &
  \begin{tabular}[c]{@{}c@{}}1,264\\6,434\end{tabular} &
  \begin{tabular}[c]{@{}c@{}}3,342\\ 183\end{tabular} &
  \begin{tabular}[c]{@{}c@{}}5,260\\ 217\end{tabular} &
  \begin{tabular}[c]{@{}c@{}}2,269\\ 169\end{tabular} &
  \begin{tabular}[c]{@{}c@{}}5,722\\ 419\end{tabular} &
  \begin{tabular}[c]{@{}c@{}}7,513\\ 167\end{tabular} &
  \begin{tabular}[c]{@{}c@{}}2,811\\ 212\end{tabular} &
  \begin{tabular}[c]{@{}c@{}}2,010\\ 213\end{tabular} &
  \begin{tabular}[c]{@{}c@{}}4,196\\ 205\end{tabular} &
  \begin{tabular}[c]{@{}c@{}}2,659\\ 400\end{tabular} &
  \begin{tabular}[c]{@{}c@{}}2,296\\ 397\end{tabular} \\ \hline
\end{tabular}%
}
\caption{Six scenarios from eleven DNN inference models, each with varying request rates (requests/s) and latencies (ms).}
\label{tab:scenarios}
\end{table*}

This section evaluates the performance of ParvaGPU in comparison to existing studies, using various scenarios listed in Table~\ref{tab:scenarios}.

\subsection{Experimental Environment}
For the experiments, we utilized multiple instances of Amazon p4de.24xlarge~\cite{EC2_P4_instance}, each equipped with eight A100 GPUs, each with 80GB of memory. Each instance has 96 vCPUs and a total of 1,152G of main memory. Regarding the software configuration, Ubuntu 20.04, CUDA 12.0.1, and PyTorch 1.14.0 were used, and all DNN inference models were sourced from NVIDIA-provided PyTorch models.

In this study, we selected existing frameworks such as gpulet~\cite{choi2022serving}, iGniter~\cite{xu2022igniter}, and MIG-serving~\cite{tan2021serving} as baselines to compare with ParvaGPU. MIG-serving offers a choice between a fast or slow algorithm, but the latter requires about 6 hours per scheduling, rendering it unsuitable for environments with fluctuating request rates. Consequently, only the fast algorithm is considered for comparison in this research. To ascertain the efficacy of using MPS in ParvaGPU, we developed ParvaGPU-single, a variant of ParvaGPU that does not employ MPS. Additionally, to determine the efficiency of the Allocation Optimization algorithm in ParvaGPU, we created ParvaGPU-unoptimized, which activates MPS but omits the final stage of optimization. These variants are not used in all experiments but only when deemed relevant.

Table~\ref{tab:scenarios} outlines six scenarios, each comprising combinations of varying request rates and latencies using 11 representative DNN inference models or services. The scale of the request rate or latency between models in each scenario was determined by referencing prior research~\cite{choi2022serving, xu2022igniter}, or by reflecting the throughput and execution times observed between models in the actual profiling results. Similar to previous studies such as gpulet and iGniter, ParvaGPU takes into consideration the queuing time of requests on the server. Consequently, the internal latency within the algorithm is set to half of the target latency~\cite{shen2019nexus} indicated for each service in Table~\ref{tab:scenarios}. Scenario 1 is designed to observe performance changes when the number of services is reduced, using six models from Scenario 2. Scenarios 2 to 6 assume various situations with a gradual increase in GPU operations. \cheolho{Scenarios 3 and 4 explore increasing request rates while maintaining the same SLO latency. Scenarios 5 and 6 reflect conditions that require high computational power, with stricter SLO latency or higher request rates.}

\subsection{Effectiveness of Spatial GPU Sharing in ParvaGPU}

In this section, the overall performance of ParvaGPU is compared and analyzed using six different scenarios.

\subsubsection{Total Number of GPUs}

\begin{figure}[t]
    \centering    
    \includegraphics[width=0.47\textwidth]{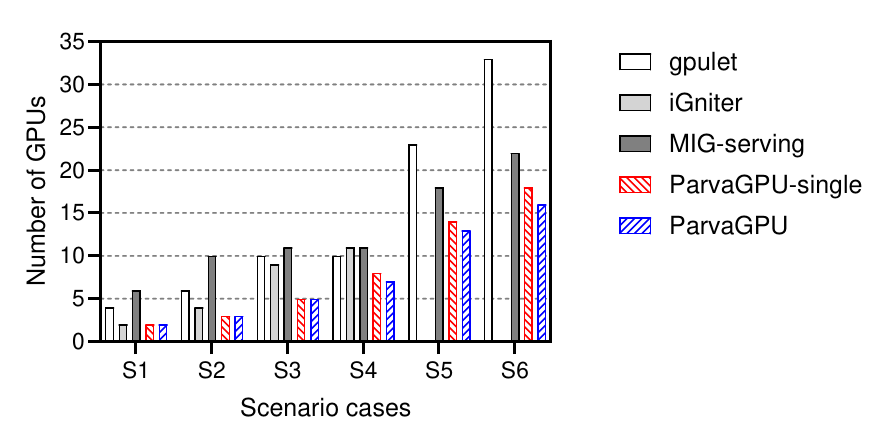}
    \caption{Total number of GPUs of each baseline and ParvaGPU.}
    \label{fig:real world num gpu}
\end{figure}

Figure~\ref{fig:real world num gpu} illustrates the number of GPUs used by each baseline and by ParvaGPU in various scenarios. Compared to gpulet, iGniter, and MIG-serving, ParvaGPU conserves an average of 46.5\%, 34.6\%, and 41.0\% in GPU usage, respectively. This indicates that ParvaGPU efficiently allocates the minimal necessary GPU resources to each service in any scenario environment, facilitated by its Segment Configurator and Segment Allocator. In S5 and S6, which are characterized by high request rates, gpulet sees a marked increase in the number of GPUs allocated. As the request rate grows, gpulet divides a service into multiple partitions. However, since only two partitions can be placed on each GPU, the usage of GPUs escalates significantly. For iGniter, while it utilizes fewer GPUs in S1 and S2, the need for more GPUs arises as the request rate increases. This increase is attributable to the limitations of its predictive model, leading to a rise in internal slack and external fragmentation. As described in the iGniter paper, iGniter is unable to manage high request rates, leading to its failure to execute in S5 and S6. MIG-serving consumes the most GPUs in scenarios with low request rates. This is due to overallocation resulting from its heuristic algorithm in scenarios with smaller request rates, as explained in Section~\ref{Internal Slack & External Fragmentation}. ParvaGPU and its variant without MPS, ParvaGPU-single, use the same number of GPUs in scenarios S1-S3, which require up to three GPUs. However, in scenarios S4, S5, and S6, where a higher number of GPUs is necessary, ParvaGPU shows a reduction of 12.5\%, 7.1\%, and 11.1\%, respectively. \cheolho{This demonstrates that ParvaGPU can further reduce costs by the same percentages compared to ParvaGPU-single when purchasing or utilizing cloud GPUs.}

\subsubsection{Internal Slack \& External Fragmentation}
\label{Internal Slack & External Fragmentation}
\begin{figure}[t]
    \centering    
    \includegraphics[width=0.47\textwidth]{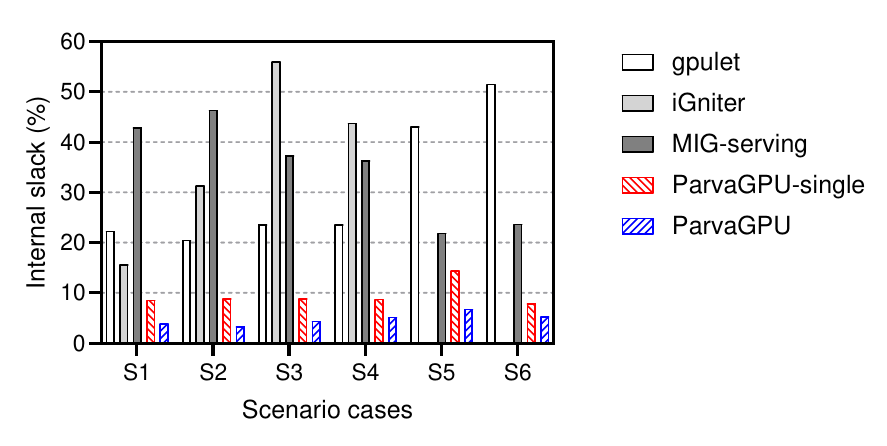}
    \caption{Internal slack rate of each baseline and ParvaGPU.}
    \label{fig:real world internal slack}
\end{figure}

We define the metric for GPU internal slack as the difference between 1 and the total SM activity rate of the GPUs, calculated using Equation~\ref{eqn:Total Internal Slack}. SM activity provides a measure of GPU utilization that reflects both spatial and temporal aspects. If a GPU has $M$ SMs, then a kernel that uses $M$ blocks throughout the time interval will produce an activity of 1 (100\%). In contrast, a kernel employing $M/5$ blocks for the same duration, or one using $M$ blocks but only active for one-fifth of the time with the SMs idle otherwise, both will have an activity of 0.2 (20\%)~\cite{nvidia_DCGM}. In Equation~\ref{eqn:Total Internal Slack}, $N$ represents the total number of services, $SM_i$ denotes the number of SMs allocated to the $i$th service and $A_i$ indicates the SM activity of the $i$th service.

\begin{equation}\label{eqn:Total Internal Slack}
\text{GPU Internal Slack} = 1 - \frac{\sum_{i=1}^{N} (SM_i \cdot A_i)}{\sum_{i=1}^{N} SM_i}
\end{equation}

Figure~\ref{fig:real world internal slack} shows the degree of GPU internal slack for each baseline in various scenarios. Compared to ParvaGPU, gpulet, iGniter, MIG-serving, and ParvaGPU-single exhibit, on average, 26\%, 32\%, 30\%, and 4.7\% more internal slack, respectively. ParvaGPU effectively finds the equilibrium point of maximum performance improvement resulting from the interaction of three factors: the size of the MIG instance, the batch size, and the number of MPS processes. These factors are crucial for minimizing internal slack. In the context of executing DNN inference models, achieving an SM activity of 1 is challenging due to factors such as data transfers between the host and GPU memory. The fact that ParvaGPU's internal slack is in the range of 3-5\% indicates that it is optimally configured to prevent internal slack. iGniter bases its resource allocations on sampling-based profiling, but accuracy limitations lead to internal slack. The analysis of other baselines will be conducted in conjunction with the analysis of external fragmentation.

\begin{figure}[t]
    \centering    
    \includegraphics[width=0.47\textwidth]{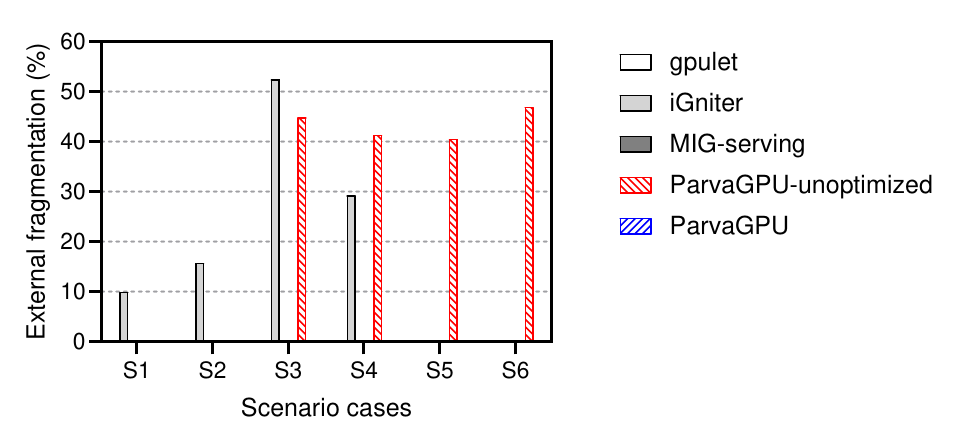}
    \caption{External fragmentation rate of each baseline and ParvaGPU.}
    \label{fig:real world external fragmentation}
\end{figure}

The degree of GPU external fragmentation is defined as follows:

\begin{equation}
\text{GPU External Fragmentation} = \frac{\sum_{i=1}^{N} (SM_i)}{G \times S}
\end{equation}
, where $N$ represents the total number of services, $SM_i$ denotes the number of SMs allocated to the $i$th service, $G$ denotes the total number of GPUs, and $S$ indicates the number of SMs contained in a single GPU.

Figure~\ref{fig:real world external fragmentation} illustrates the degree of GPU external fragmentation for each framework in different scenarios, highlighting that ParvaGPU completely eliminates external fragmentation in all scenarios. In particular, ParvaGPU reduces external fragmentation by an average of 29\% compared to ParvaGPU-unoptimized, demonstrating the effectiveness of ParvaGPU's Allocator Optimization algorithm in eliminating external fragmentation. gpulet avoids external fragmentation by allocating the remaining GPU resources entirely to the second partition when assigning two partitions to one GPU. However, as shown in Figure~\ref{fig:real world internal slack}, it does not account for the internal slack of this partition. iGniter, lacking a mechanism to resolve external fragmentation, experiences an average of 26.9\% external fragmentation. MIG-serving prevents external fragmentation by scoring configurations, where configurations with external fragmentation are scored lower and thus not selected. However, even in scenarios with low request rates, MIG-serving prioritizes preventing external fragmentation, leading to resource overallocation and resulting in internal slack, as shown in Figure~\ref{fig:real world internal slack}.

\subsection{Evaluation of Inference Server Services' Quality}

\subsubsection{SLO Violation Rate}

\begin{figure}[t]
    \centering    
    \includegraphics[width=0.47\textwidth]{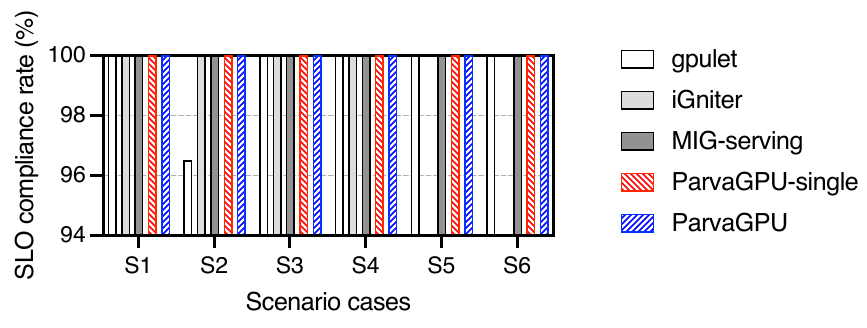}
    \caption{\cheolho{SLO compliance rate of each baseline and ParvaGPU.}}
    \label{fig:slo violation}

\end{figure}

An SLO violation occurs when an inference server performing DNN inference fails to meet the established SLO, which includes latency and throughput metrics, for requests processed in each step according to the batch size handled. To assess this occurrence, the proportion of batches that did not satisfy the SLO during the entire batch execution is measured and compared. \cheolho{Figure~\ref{fig:slo violation} shows the SLO compliance rate, and the SLO violation rate can be calculated as 1 minus the SLO compliance rate.} The majority of frameworks did not experience any SLO violations. However, gpulet encountered a 3.5\% rate of SLO violations in scenario 2. This can presumably be attributed to inaccuracies in gpulet's interference estimation. \cheolho{iGniter is unable to handle high request rates, so the results for scenarios S5 and S6 are not shown.}

\subsubsection{Scheduling Delay}
\label{Scheduling Delay}

\begin{figure}[t]
    \centering
    \includegraphics[width=0.47\textwidth]{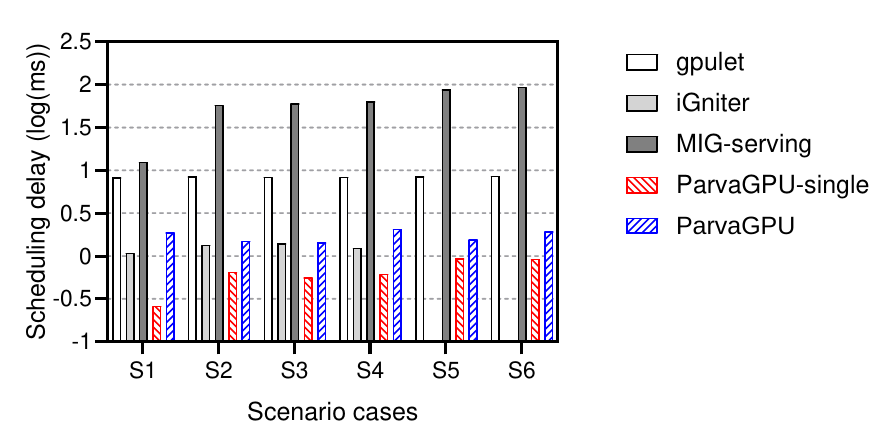}
    \caption{Scheduling delay of each baseline and ParvaGPU.}
    \label{fig:real world overhead}
\end{figure}

ParvaGPU exhibits exceptional efficiency as an inference server in terms of scheduling delay when compared to each baseline algorithm. As illustrated in Figure~\ref{fig:real world overhead}, ParvaGPU records scheduling delays that are, on average, 80\% and 97.2\% lower than gpulet and MIG-serving, respectively. This difference becomes even more pronounced in scenarios where the request rate increases. This significant reduction in overall scheduling delay is the result of ParvaGPU addressing the problem through a two-stage approach, involving a Configurator and an Allocator, and by effectively reducing the time complexity with the Demand Matching algorithm. MIG-serving performs both the determination of instance size and the allocation of resources within a large search space without separation. It considers all possible configurations of services that can be deployed in MIG configurations, which leads to increased execution times in scenarios with high request rates and numerous services. gpulet considers only two services for execution on each GPU, which leads to reduced time consumption and similar durations across the majority of scenarios compared to MIG-serving. iGniter estimates the necessary resource allocations based on sparse profiling values, thus taking approximately 35\% less time than ParvaGPU. However, such estimates unavoidably result in internal slack, leading to inefficient outcomes with respect to the total number of GPUs. Compared to ParvaGPU-single, ParvaGPU incurs an additional 1.1ms of scheduling delay, due to ParvaGPU-single not exploring various scenarios based on the number of processes, thereby achieving faster scheduling.

\subsection{Model Scalability Evaluation with Predictor}

Each baseline and ParvaGPU offer a predictor to facilitate the design of GPU deployments, even in the absence of actual GPUs. Using this predictor, we incrementally increase the number of services in S5, characterized by a high request rate, and measure both the number of GPUs utilized and the scheduling delay in each framework. This experiment simulates scenarios where a client intends to significantly expand their service offerings or when a cloud provider delivers inference services, aiming to host multiple models on a single infrastructure setup. iGniter has been excluded from this experiment due to its incompatibility with S5.

\subsubsection{Total Number of GPUs}
\label{model_scale_Total Number of GPUs}

\begin{figure}[t]
    \centering    
    \includegraphics[width=0.47\textwidth]{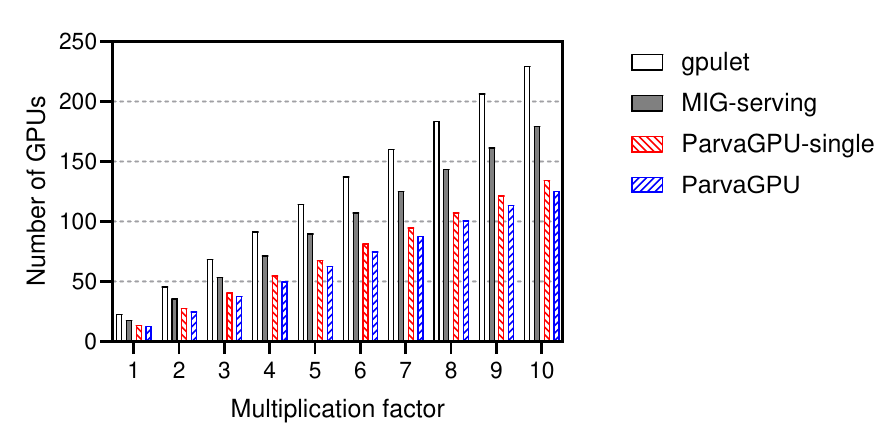}
    \caption{Total number of GPUs of each baseline and ParvaGPU with an increasing number of services in S5 from 1 to 10 fold.}
    \label{figure:num service num gpu}
\end{figure}

Figure~\ref{figure:num service num gpu} compares the number of GPUs required as the number of services in S5 increases from 1 to 10 fold. ParvaGPU uses on average 45.2\%, 30\%, and 7.4\% fewer GPUs compared to gpulet, MIG-serving, and ParvaGPU-single, respectively. This demonstrates ParvaGPU's efficiency in minimizing internal slack and external fragmentation, even as it processes a greater number of services. In the case of other baselines, they exhibit a performance pattern similar to that shown in Figure~\ref{fig:real world num gpu}.

\subsubsection{Scheduling Delay}

\begin{figure}[t]
    \centering    
    \includegraphics[width=0.47\textwidth]{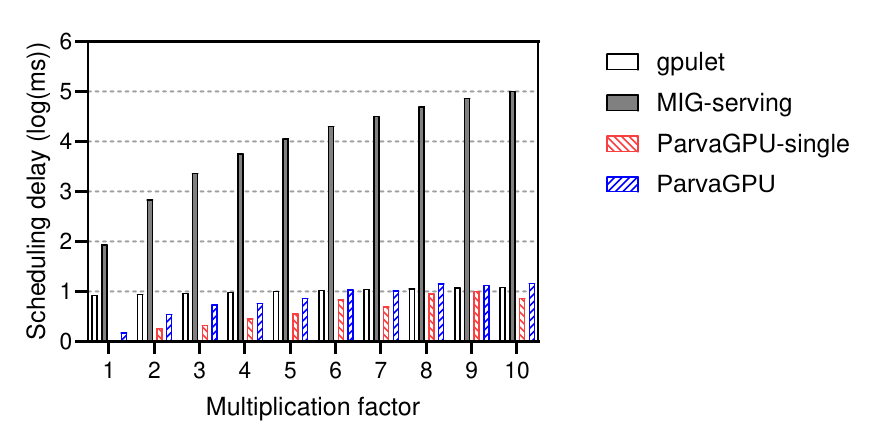}
    \caption{Scheduling delay of each baseline and ParvaGPU with an increasing number of services in S5 from 1 to 10 fold.}
    \label{figure:num service scheduling time}
\end{figure}

Figure~\ref{figure:num service scheduling time} compares the scheduling delay across frameworks as the number of services in S5 increases from 1 to 10 fold. ParvaGPU was able to reduce the delay by on average 15.8\% and 99.9\% compared to gpulet and MIG-serving, respectively. However, due to its exploration of MPS process counts, ParvaGPU experienced a slight increase in delay compared to ParvaGPU-single. MIG-serving, despite proposing a fast algorithm based on a greedy algorithm, incurs a significant scheduling overhead due to inefficient use of the search space as the number of services increases.

\section{Discussion} \label{Discussion}
\cheolho{This section discusses the applicability of ParvaGPU to various GPU architectures and examines the impact of memory-intensive models on spatial GPU sharing.

\textbf{Applicability to non-NVIDIA GPUs and recent NVIDIA architectures:} NVIDIA’s MIG feature is a cutting-edge technology not yet available on non-NVIDIA GPUs. Currently, ParvaGPU’s algorithms require GPUs that support fully isolated instance partitioning, such as MIG. However, non-NVIDIA GPUs and NVIDIA technologies tend to converge over time. For example, AMD’s compute unit masking~\cite{chow2023krisp} offers functionality similar to NVIDIA’s MPS. Should non-NVIDIA GPUs support fully isolated instance partitioning in the future, ParvaGPU’s algorithms can be adapted to the new architecture with minimal modifications. All NVIDIA GPUs adopting MIG across the Ampere, Hopper, and latest Blackwell architectures maintain identical MIG configurations. Therefore, all ParvaGPU algorithms can generally be applied to the new generations of NVIDIA GPUs.

\textbf{Impact of memory-intensive models on spatial GPU sharing:}
As Large Language Models (LLMs) and generative AI models grow in parameter size, the demand for GPU memory increases, reducing the feasibility of utilizing smaller GPU segments in ParvaGPU. However, for inference, there is growing research focused on developing smaller models. For example, a lightweight LLaMA model~\cite{touvron2023llama} requires only 7GB of memory while maintaining accuracy close to that of larger models. Furthermore, applying QLoRA tuning to the Guanaco model~\cite{dettmers2024qlora} results in memory usage of 5GB for 7B parameters and 41GB for 65B parameters. Given that NVIDIA’s H200 GPU with MIG offers 141GB and the B200 GPU provides 192GB of GPU memory, there remains potential for spatial GPU sharing, even for LLMs and generative models.
}

\section{Conclusion} \label{Conclusion}
In cloud environments, GPU-based DNN inference servers must satisfy the SLO latency for each workload under a given request rate while also minimizing GPU resource consumption. However, previous studies have not fully met these objectives. In this paper, we proposed ParvaGPU, a spatial GPU sharing technique that combines MIG and MPS technologies for high inference throughput of various DNN models in cloud environments. ParvaGPU significantly reduces GPU usage by entirely minimizing underutilization within allocated GPU space partitions and external fragmentation.

\section*{Acknowledgment}
This work was supported by the Electronics and Telecommunications Research Institute grant funded by the Korean government [23zs1300, Research on High Performance Computing Technology to overcome limitations of AI processing]. This work is also based on work supported by the National Science Foundation (NSF) under Grants No.\ 2315851,2106634, a Sony Faculty Innovation Award (Contract AG3ZURVF) and a Cisco Research Award (Contract 878201).

\bibliographystyle{IEEEtran}
\bibliography{IEEEabrv,ref}

\end{document}